\documentclass{WileyMSP-template}
\usepackage{graphicx}

\usepackage{amssymb}
\usepackage{amsmath}
\usepackage{upgreek}
\usepackage{lineno}
\usepackage{gensymb}
\usepackage[utf8]{inputenc}
\usepackage[dvipsnames]{xcolor}

\begin{document}

\pagestyle{fancy}
\setlength{\headheight}{24.81955pt}
\rhead{\includegraphics[width=2.5cm]{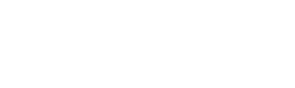}}

\color{red}
\textbf{This is the accepted version of the article published open acces in Small Methods:}

https://dx.doi.org/10.1002/smtd.202400034
\color{black}
\title{Support-based transfer and contacting of individual nanomaterials for in-situ nanoscale investigations}

\maketitle

\author{Simon Hettler*}
\author{Mohammad Furqan}
\author{Raul Arenal*}

\begin{affiliations}
Dr. S. Hettler, M. Furqan\\
Laboratorio de Microscopías Avanzadas (LMA), Universidad de Zaragoza, 50018 Zaragoza, Spain\\
Instituto de Nanociencia y Materiales de Aragón (INMA), CSIC-Universidad de Zaragoza, 50009 Zaragoza, Spain\\
Email Address: hettler@unizar.es

Dr. R. Arenal\\
Laboratorio de Microscopías Avanzadas (LMA), Universidad de Zaragoza, 50018 Zaragoza, Spain\\
Instituto de Nanociencia y Materiales de Aragón (INMA), CSIC-Universidad de Zaragoza, 50009 Zaragoza, Spain\\
ARAID Foundation, 50018 Zaragoza, Spain\\
Email Address: arenal@unizar.es

\end{affiliations}


\begin{abstract}
Although in-situ transmission electron microscopy (TEM) of nanomaterials has been gaining importance in recent years, difficulties in sample preparation have limited the number of studies on electrical properties. Here, a support-based preparation method of individual 1D and 2D materials is presented, which yields a reproducible sample transfer for electrical investigation by in-situ TEM. A mechanically rigid support grid facilitates the transfer and contacting to in-situ chips by focused ion beam with minimum damage and contamination. The transfer quality is assessed by exemplary specimens of different nanomaterials, including a monolayer of WS\textsubscript{2}. Possible studies concern the interplay between structural properties and electrical characteristics on the individual nanomaterial level as well as failure analysis under electrical current or studies of electromigration, Joule heating and related effects. The TEM measurements can be enriched by additional correlative microscopy and spectroscopy carried out on the identical object with techniques that allow a characterization with a spatial resolution in the range of a few microns. Although developed for in-situ TEM, the present transfer method is also applicable to transferring nanomaterials to similar chips for performing further studies or even for using them in potential electrical/optoelectronic/sensing devices.

\end{abstract}

\section{Introduction}

The detailed study of structure, composition and properties of nanomaterials is crucial for an in-depth understanding, to achieve an improvement of their synthesis and to yield a successful application. With respect to electrical properties of 1D nanomaterials, great effort has been invested in the characterization of numerous individual nanotubes (NTs) with the help of scanning electron microscopy (SEM) and electron-beam lithography (EBL), e.g., of carbon NTs \cite{Yao.2000,Huang.2015} but also of related structures \cite{Brunbauer.2016,Lin.2018,Empante.2019,Roy.2022}. This approach allows the analysis of electrical properties of individual nanomaterials, but the exact structure and composition of that particular nanomaterial, including defects or dopants down to the atomic level, is typically not acquired. Thus, such studies only give ensemble values for the, e.g., conductivity and lack the determination of the structure-property relation at the individual nanomaterial level. 

Transmission electron microscopy (TEM) and related spectroscopic techniques are frequently used to study the chemical composition and the structure of individual nanomaterials even at atomic resolution. In recent years, in-situ or operando TEM studies have been gaining importance to investigate nanomaterials under application-relevant conditions and to trace and understand their evolution under external stimuli. In comparison to in-situ heating studies or experiments in gas or liquid environments, any electrical measurement requires an additional, controlled contacting of the sample. Therefore, careful preparation and contacting has been necessary for in-situ biasing studies \cite{Sato.2017,Zhang.2017,MolinaLuna.2018,Ishida.2020,Arita.2020,Luong.2020,Nukala.2021}. The preparation is mostly performed with the help of the focused ion beam (FIB) technique \cite{Sato.2017,Zhang.2017,Gorji.2020,Nukala.2021} or by sample preparation directly on the in-situ TEM chip \cite{denHertog.2012,Kozlova.2013,MolinaLuna.2018,Luong.2020,Hsueh.2023}. Most of these studies are based on bulk materials, where the final thinning step of the sample can be performed on the chip and after contacting, allowing to clean previously deposited Pt-based contamination \cite{MartialDuchamp.2014,Vijayan.2017,Sato.2017,Nukala.2021}. In contrast, the inevitable damage and contamination of the specimen when trying to use a FIB to directly transfer and contact, e.g., an individual NT, has limited the number of studies, where individual nanomaterials have been transferred and used as specimen for in-situ TEM electrical characterization \cite{Xu.2011,Hettler.2021}. Some electrical in-situ studies have been conducted with the help of a scanning tunneling microscopy tip (Nanofactory\textsuperscript{TM} TEM holder)\cite{Golberg.2006,Aslam.2011,Arenal.2011}, but the approach has strong requirements for sample geometry, lacks reproducibility and electrical contacting is challenging.  Similar to SEM-based studies, EBL can be used to contact nanomaterials dispersed on a thin, electron-transparent membrane for in-situ electrical TEM \cite{denHertog.2012,Luong.2020,Hsueh.2023}. However, in the drop-casting process, material can pile up on the contacts of the in-situ chip and furthermore, the specimen should be suspended in vacuum for optimum TEM conditions.   A FIB-based sample preparation of individual nanoparticles suspended on a thin film has been suggested for conventional TEM and electron tomography analysis \cite{Huang.2022}.

In this paper, we present a support-based method that allows the preparation of a previously-studied 1D or 2D nanomaterial for in-situ electrical TEM studies with minimum contamination and damage. The method allows to screen a conventional TEM sample for a suitable or desired individual 1D or 2D material and subsequently transfer it in a FIB-assisted process to an in-situ chip. The transfer requires the mechanical stability of a silicon nitride based TEM grid, which also strongly facilitates the contacting of the material to the contact pads. Moreover, this approach allows correlative microscopy, e.g., Raman measurements before and after the in-situ TEM experiment. The transfer method has been developed for in-situ TEM but can be used to transfer nanomaterials to chips with similar design. In this manuscript, a general guideline of the preparation method is explained in detail and the transfer quality is evaluated for different samples.

\section{Support-based nanomaterial transfer}
\label{S:SamplePrep}

\subsection{Process description}
\label{S31}

\begin{figure}[t]
    \centering
    \includegraphics[width=0.7\linewidth]{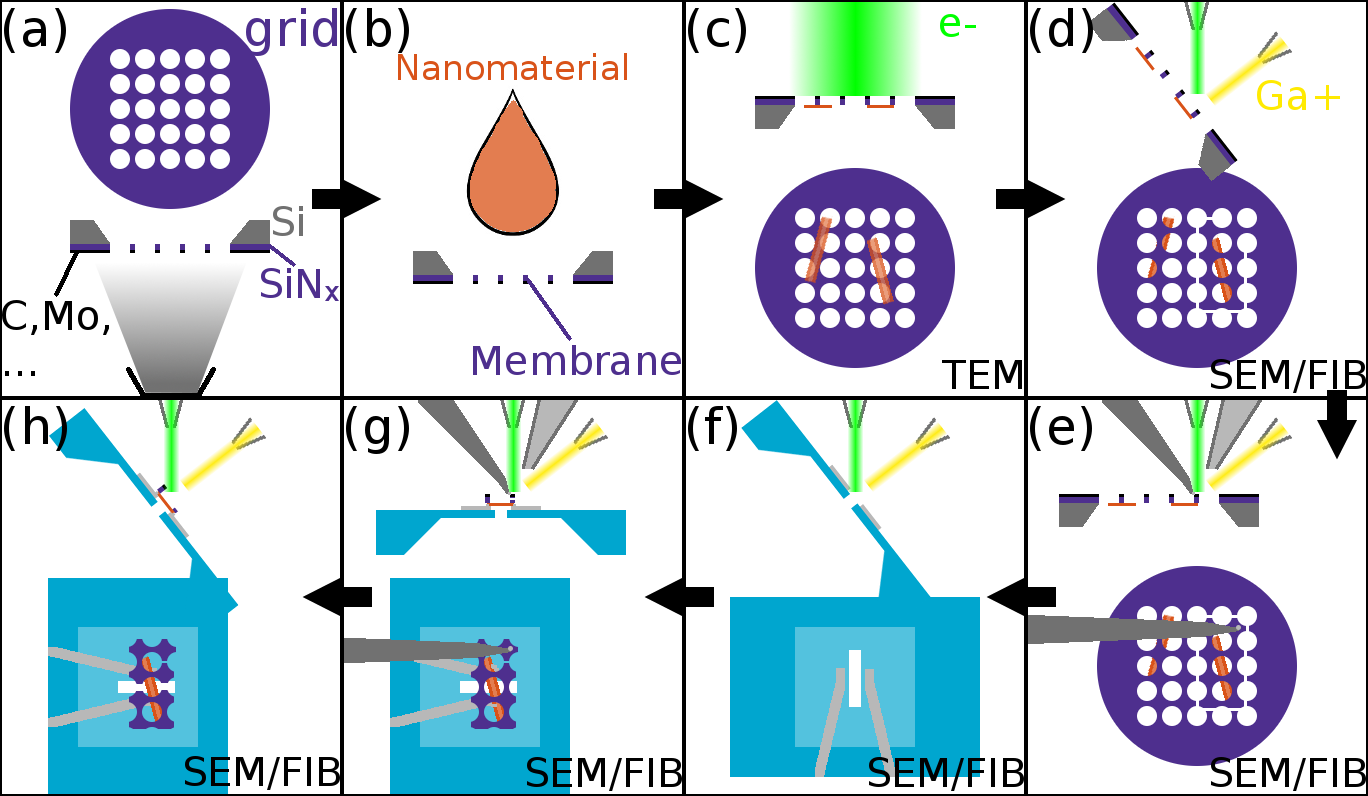}
    \caption{ Sketch of the transfer process, which starts with (a) a holey SiN\textsubscript{x} TEM grid coated from the front side with 10~nm of, e.g., amorphous carbon (step 1). (b) The nanomaterial is subsequently deposited on the back side by, e.g., drop-casting (step 2). (c) One or several suitable nanomaterials are selected in a TEM analysis of the sample (steps 3+4). (d) The membrane containing the selected nanomaterial is cut by FIB at a sample tilt of 52\degree, leaving a bridge to the surrounding membrane (steps 4+5). (e) The nanomaterial sustained by the membrane is lifted out using a micro needle and FIBID followed by FIB cutting of the remaining bridge (steps 6+7). (f) A hole is milled between the contacts of an in-situ chip (step 8). (g) The membrane + nanomaterial is put in contact with the chips' contact pads with the micro needle and a FIBID process (step 9-11). (h) The SiN\textsubscript{x} membrane in the hole area is removed by FIB, leaving the nanomaterial as only connection between the pads of the chip (step 12).}
    \label{F:FigSketch}
\end{figure}

 The key element of the transfer process is the use of a holey TEM grid made of a mechanically rigid silicon nitride (SiN\textsubscript{x}) membrane. After a metal coating of the grid, the nanomaterial is deposited on the grid which allows a pre-screening by TEM to select a specific individual nanomaterial. The transfer itself is conducted in a dual-beam instrument SEM/FIB with micro needle and a gas-injection system (GIS) facilitating focused ion beam induced deposition (FIBID). The following list gives a point-by-point description of the transfer process sketched in \textbf{Figure~\ref{F:FigSketch}}, FIB conditions are discussed in the following section. Figure~\ref{F:FigSketch}a and b introduce the denomination of important elements of the processes: With the term "grid" we refer to the whole TEM grid with 3~mm diameter including the Si frame, while "membrane" denotes only the thin, 200~nm thick SiN\textsubscript{x} membrane in the center of the grid.

\begin{figure}[t]
    \centering
    \includegraphics[width=0.9\linewidth]{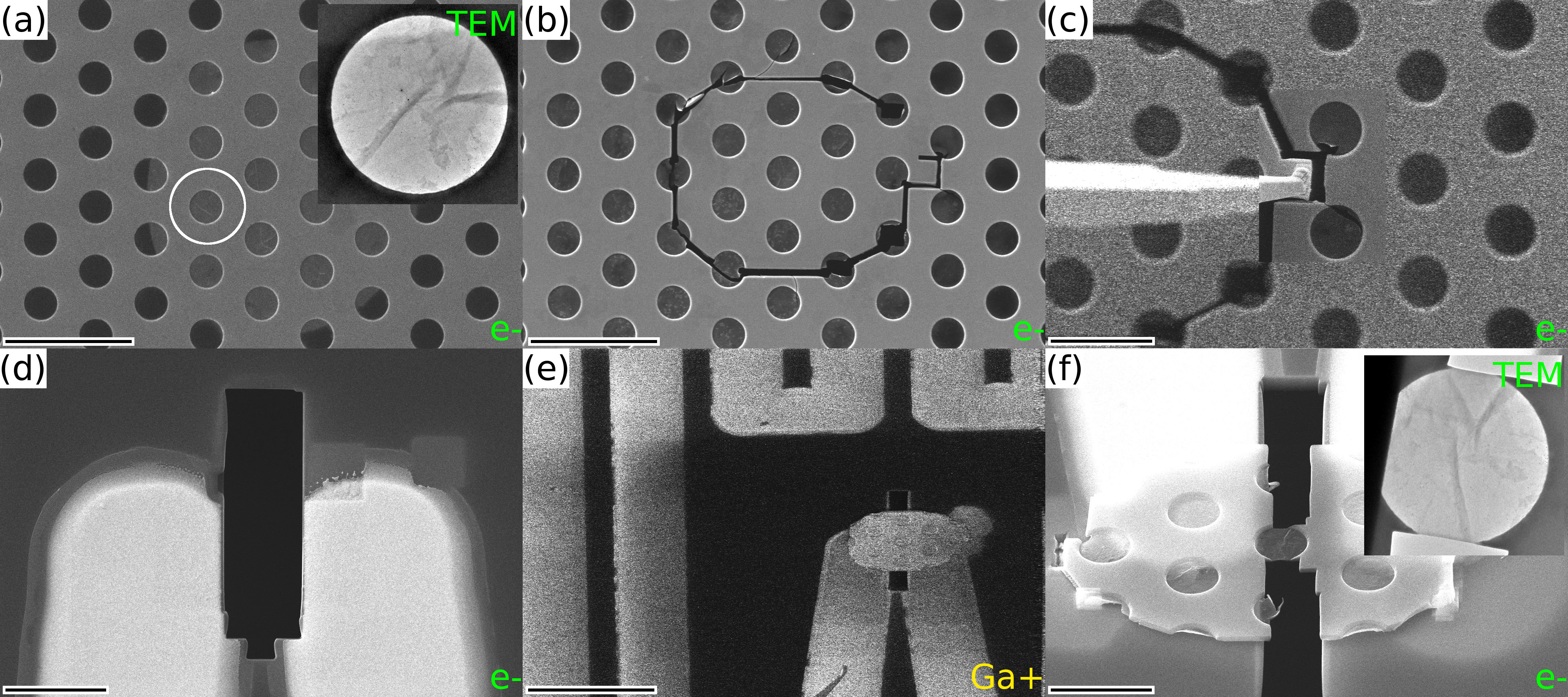}
    \caption{SEM/FIB documentation of exemplary sample preparation. (a) SEM image of a GO film  suspended on the lower side of the SiN\textsubscript{x} membrane of a carbon-coated holey TEM grid. Inset shows a TEM image (image width 3~$\upmu$m) of sample position marked by a white circle (step 3). (b) SEM image of the sample after cutting the membrane with remaining bridge and landing zone (step 5). (c) Image composed of two subsequent scans showing micro needle holding membrane and nanomaterial after contacting and cutting free (step 7). (d) SEM image of in-situ chip with hole milled between contacts (step 8). (e) FIB image of membrane and nanomaterial contacted to the in-situ chip with the needle retracted (step 11). (f) SEM image after milling of the membrane in the hole area of the chip with nanomaterial forming the only bridge between contacts. Inset shows a TEM image (image width 3~$\upmu$m) after transfer (step 12). Scale bars are (a,b) 10~$\upmu$m, (c,d,f) 5~$\upmu$m, (e) 20~$\upmu$m.}
    \label{F:GOtrans}
\end{figure}

\textbf{Figure~\ref{F:GOtrans}} shows six SEM or FIB images covering the FIB part of the process for an example transfer using a graphene oxide (GO) sample \cite{GO_SynthVictorRoman.2020}. In the drop-casting process of the GO, individual flakes form a continuous film on the TEM grid. In the following, "flake" refers to an individual single-crystalline sheet and "film" describes a continuous sheet irrespective of its crystalline structure. Figure~S1 in the supporting information (SI) shows a second exemplary transfer of a NT and flake made of the misfit-layered compound (MLC) LaS-TaS\textsubscript{2} \cite{Lajaunie.2018,Hettler.2020}.

\vspace{0.3cm}
\textit{Description of transfer procedure:}

\begin{enumerate}
    \item Thin (10-20~nm) metal coating of the holey TEM grid from the front side, e.g., with carbon or molybdenum (Figure~\ref{F:FigSketch}a). The coating is required to minimize charging of the SiN\textsubscript{x} in both TEM and SEM/FIB. Additionally, it facilitates a rapid FIBID process on the coated surface.  
    \item Conventional TEM sample preparation (e.g. drop casting) on the back side of the coated grid (Figure~\ref{F:FigSketch}b). For the example in Figure~\ref{F:GOtrans}, GO was deposited on the back side of a SiN\textsubscript{x} grid, previously coated on the top side with Mo.   The reason for the preferable front-side coating and back-side deposition is due to a geometrical constraint discussed below. A ligand-cleaning step in an activated carbon + ethyl alcohol bath can be performed at this stage \cite{Li.2021}.
    \item A TEM analysis of the sample, including electron spectroscopic techniques (energy-dispersive X-ray spectroscopy (EDX) and electron energy-loss spectroscopy (EELS)) allows the selection of a nanomaterial with desired characteristics (Figure~\ref{F:FigSketch}c and inset in Figure~\ref{F:GOtrans}a). The acquisition of a low-magnification image allows to mark the location of the nanomaterial to facilitate finding its position in SEM or additional correlative microscopy techniques that can be performed on the individual nanomaterial basis before the specimen transfer.
    \item  Mounting the TEM grid in the dual-beam instrument with the coated front side pointing upwards (Figure~\ref{F:FigSketch}d and \ref{F:GOtrans}a). As the nanomaterial is located on the bottom side of the membrane, the insulating SiN\textsubscript{x} layer between nanomaterial and contact pad is avoided after transfer to the chip, crucial for homogeneous electrical contact of 2D materials as discussed below. Identification of the hole with selected nanomaterial in SEM (Figure~\ref{F:GOtrans}a).  
    \item Cut of SiN\textsubscript{x} membrane around the hole containing the nanomaterial by FIB (stage tilted to 52\degree) leaving one remaining bridge and preparation of "landing zone" for take-out needle (Figure~\ref{F:FigSketch}d and \ref{F:GOtrans}b).  Although the SiN\textsubscript{x} is "low-stress", it can crack upon milling. When milling inside the holey area, these cracks remain local and do not affect the overall stability of the membrane. However, when milling between two holes at the edge of the hole array, the risk is high that the whole membrane breaks. Therefore, the outer holes of the array should be avoided.  
    \item Contacting take-out needle to landing zone by FIBID of Pt/C with stage tilted to 0\degree ~(Figure~\ref{F:FigSketch}e). 
    \item  After a waiting time to allow for desorption of adsorbed precursor molecules from the nanomaterial (we first waited until vacuum was better than 2$\cdot$10\textsuperscript{-6}~mbar and then waited an additional 10~min), the remaining bridge is cut open by FIB and the needle with attached SiN\textsubscript{x} membrane carrying the selected nanomaterial is retracted (Figure~\ref{F:FigSketch}e and \ref{F:GOtrans}c). 
    \item Milling of a hole between the contacts of an in-situ chip with dimensions according to the selected nanomaterial and holey SiN\textsubscript{x} membrane, typically with a width between 1 and 4~$\upmu$m and a length of more than 10~$\upmu$m (Figure~\ref{F:FigSketch}f and \ref{F:GOtrans}d). This step is performed with the stage tilted to 52\degree  and with the FIB operating at 30~kV and a current of 0.79 nA, which is decreased to 80 pA for final polishing of the edges of the hole.
    \item Transfer of the sample to the in-situ chip: Careful approach of membrane and nanomaterial to the metal contacts (Figure~\ref{F:FigSketch}g). Due to electrostatic charging, the SiN\textsubscript{x} membrane attaches easily to the metal contact pads.
    \item Contacting of nanomaterial to metal contacts by Pt-FIBID (Figure~\ref{F:FigSketch}g). This deposition ensures both mechanical fixation and electrical contacting. As discussed below, this is a crucial aspect as the quality of the contacts is critical for mechanical fixation and for performing reliable electrical measurements, but, on the other hand, extensive FIBID inflicts contamination to the nanomaterial. Thus, a careful balance should be sought when dealing with those aspects that compromise the quality of the sample and the contacts. 
    \item  Waiting for desorption of adsorbed precursor molecules from the nanomaterial (similar to step 7),   followed by disconnection of landing zone with attached micro needle from rest of SiN\textsubscript{x} membrane by FIB milling and retraction of micro needle (Figure~\ref{F:GOtrans}e). 
    \item Removal of SiN\textsubscript{x} in the hole area of the in-situ chip by FIB milling (stage tilted to 52\degree) in order that the nanomaterial is the only connection between the contact pads (Figure~\ref{F:FigSketch}h and \ref{F:GOtrans}f).
   \end{enumerate}
   
    After this last step, an in-situ TEM experiment and possible correlative microscopy/spectroscopy measurements can be conducted.

The list describes the process with the nanomaterial deposited on the back side and metal coating of the front side. This orientation is preferable due to the following reason: To guarantee a good electrical contact on the final device, the nanomaterial should be in direct contact with the pads of the in-situ chip to avoid the insulating barrier of the SiN\textsubscript{x} membrane. This could also be achieved in an inverted setup with a back-side coating, a front-side deposition of the nanomaterial and mounting the TEM grid upside down in the dual-beam instrument. However, the membrane is then located 200~$\upmu$m (thickness of the Si frame of the grid) below the grid surface. This causes a potential crash with the GIS needle, which typically has a smaller distance to the surface and, additionally, the FIB view is limited due to the inclined incidence of the beam. Nevertheless, the upside-down configuration is in general possible and could be used for nanomaterials, which are difficult to prepare on the back side of the TEM grid.

Regarding the electrical contacting of the nanomaterial, differences exist between 1D and 2D materials. For 2D materials, it is imperative to have a direct, planar connection to the contact pad of the in-situ chip. In this case, FIBID is required only at a few points to mechanically fix the membrane on the grid as electrical contact is generated by direct contact. In contrast, if the insulating SiN\textsubscript{x} is located between the 2D material and the pad, many FIBID deposits are required at the edge of the membrane to bridge the insulating barrier and to guarantee a homogeneous contact and current feed. As FIBID is the major source of contamination, it would have a negative impact on the sample quality. On the other hand, for 1D materials, the location of the membrane with respect to pad and nanomaterial is not as critical, as a good contact between pad and 1D material is already achieved by two FIBID deposits fixing and connecting the respective ends to the pad of the in-situ chip (see Figures~\ref{F:FigEva} and \ref{F:Elec} and Figure~S1 in SI). It is noted that 1D nanomaterials can end up on both top and bottom side of the membrane in a drop-casting process.

\subsection{Minimization of damage and contamination}
\label{S32}

For the best transfer, damage and contamination of the individual nanomaterial have to be avoided. The main source of damage is a direct impact of heavy Ga\textsuperscript{+} ions on the nanomaterial. Therefore, imaging of the sample area with primary Ga\textsuperscript{+} ions has to be completely avoided in the transfer process by imaging only areas outside of the actual sample position using small scanning windows (see Figure S1b in SI).
To minimize damage by secondary ions created during milling or imaging of surrounding areas, FIB voltage and current should be chosen accordingly. The minimum values for both should be chosen, where sufficient imaging and milling quality is obtained. For the pre-cutting of the membrane (step~5), 16~kV yielded best results, as milling quality decreases and Ga implantation increases when lowering the voltage even further. Thus, in most example transfers, we used 16~kV as well for the other steps,  except for the hole milling between the contacts of the in-situ chip (step 8), which was performed at 30~kV and currents of 0.79~nA for initial opening and 80~pA for polishing of the edges.  As discussed below, the voltage was reduced to 5~kV for FIBID and subsequent milling processes (steps 6 \& 7 and 10 \& 11) for WS\textsubscript{2} monolayer samples to minimize contamination. Ion beam currents were 23 pA for step 5 and 11 pA for the remaining FIB applications in the example tranfer of the GO sample (Figure~\ref{F:GOtrans}).

In the case of 2D materials, such as the GO sample, which are suspended over the entire hole of the SiN\textsubscript{x} membrane, direct impact of the Ga\textsuperscript{+} ions is inevitable in step 12 (removal of the membrane in the central part to leave the nanomaterial as the only connection between the contacts). This impact is limited to the edges of the sample and can cause damage as well as Ga implantation, as shown and discussed in section~\ref{S:Eval}. In case of a 1D material, this source of damage and contamination can be completely avoided if the distance between the hole edge and the nanomaterial is sufficiently large.

Contamination arises from the deposition of Pt-based precursor gases adsorbed on the nanomaterial induced by secondary electrons and ions. Secondary electrons are generated both by primary electrons and ions, so imaging with both beams in areas around the nanomaterial position has to be kept at a minimum during and after a FIBID process. Imaging with the electron beam was mainly performed at 5 keV and 100 pA but the electron energy was increased to 16 keV for imaging the in-situ chip after the transfer to avoid excessive charging of its supporting membrane.  This charging can be reduced by a carbon coating of the back side of the in-situ chip, but not completely avoided due to the large thickness of the employed membrane of 1~$\upmu$m. 

In some experiments, focused electron beam induced deposition (FEBID) using 5~keV and 1.7~nA has been tried. However, FIBID has provided a higher success rate for contacting the micro needle to the SiN\textsubscript{x} membrane at minimum dose, as the ion beam incident angle (stage at 0\degree) allows to directly irradiate the area between needle and membrane. The tip of the needle was thinned down to less than a $\upmu$m and we used a FIBID deposition window with a size of approximately 0.8x0.8~$\upmu$m\textsuperscript{2} for a current of 11~pA. For a fast and effective contacting of the micro needle to the membrane, a conductive coating of the membrane surface has proven to be helpful, requiring the metal coating in step 1. A possible reason could be that the adhesion of the precursor gas molecules on the uncoated insulating SiN\textsubscript{x} is low.

As the precursor gases remain adsorbed on the sample for a prolonged time, even after closing the valve, imaging or milling has to be avoided for some time after the FIBID process. The factor that limits the waiting time is mechanical drift of the lift-out needle. A large drift and resulting mechanical pressure on the substrate may lead to a rupture of the membrane. To minimize this drift, the micro needle should be driven with a small final speed ($<$1 $\upmu$m/s) and let come to a rest for a few minutes close to the landing zone before contacting.  In the conducted experiments, we first waited for the vacuum to decrease below 2$\cdot$10\textsuperscript{-6}~mbar and then waited an additional 10 minutes. 

Another possible source of contamination is redeposition, which is caused during FIB milling when the material that is milled away adsorbs on areas close by. In the cutting step of the SiN\textsubscript{x} membrane, before take out, the milling areas are located at a rather large distance of a few microns to the actual sample area. In step 12, when milling away the membrane on the in-situ chip, cleaning cross sections with a direction towards the sample are employed. In the performed experiments, effects derived from redeposition could not be detected (see also section~\ref{S:Eval}).

\subsection{Studied alternative approaches}
\label{S33}

Several alternative approaches were studied in this work. Initial tries to transfer nanomaterials by direct contacting to the micro needle were not satisfactory due to stronger inevitable contamination during the FIBID/FEBID contacting processes. In addition to the SiN\textsubscript{x} support, holey TEM grids made of thick amorphous carbon and gold were tested. However, both films do not possess the mechanical stability necessary to completely avoid bending of the films during the transfer,    except for cases where the nanomaterial itself exhibits sufficient mechanical stiffness.   Moreover, these films usually do not attach easily on the second contact pad located away from the micro needle. For these reasons, these approaches have been discarded but remain as tentative alternatives.

\medskip
\newpage

\begin{figure}[h]
    \centering
    \includegraphics[width=0.7\linewidth]{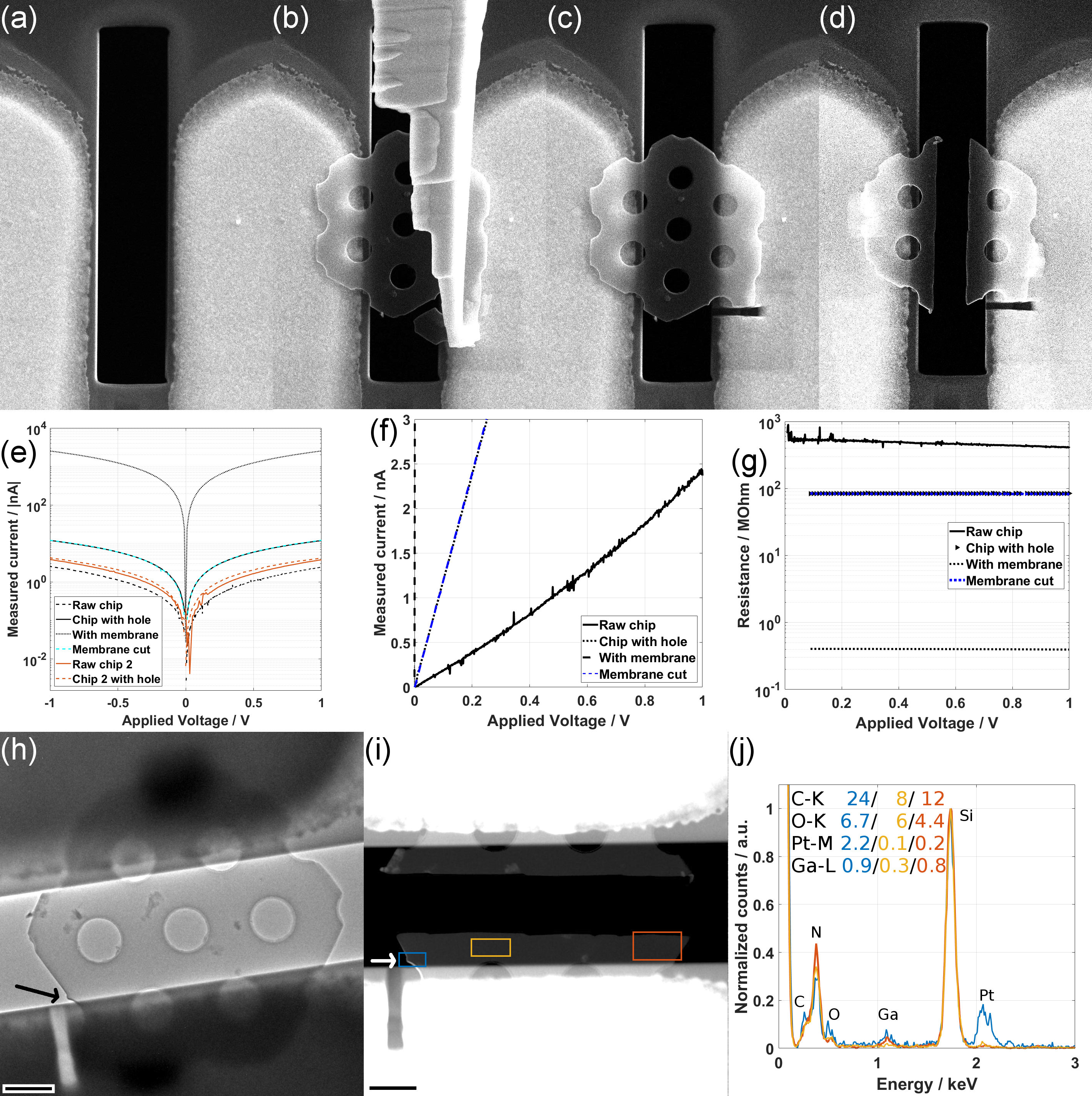}
    \caption{Study of reference sample preparation. SEM images of the contact pads of the reference chip (a) after milling the rectangular hole (step 8), (b) with the membrane in contact with the pads and micro needle still attached (step 9), (c) after cutting and removal of the micro needle (step 11) (d) after fixation of the membrane by FIBID and opening the membrane by FIB (step 12). (e) I-V characterization of the reference chip (black and blue lines as indicated) and a second chip (orange lines) at different stages. Note that curves of chip with hole (black) and chip with cut membrane (light blue dashed) coincide. (f) I-V curve at linear scale showing the non-linear component of the raw chip. The curve for chip with membrane (black dashed) falls on the y-axis. (g) Resistance determined from the I-V curves in (e). (h) TEM image of the chip with transferred membrane after step 11 shows a dark halo at the left edge close to the needle contact  (dark arrow). (i) HAADF-STEM image of the chip after step 12 shows a bright halo on the lower left edge (white arrow). (j) EDX comparison of three areas marked in (i) shows that Pt contamination is present only on one side of the chip (blue) and absent in the center (yellow) and on the opposite side (red). Composition of contamination elements is indicated in at\%.   Scale bars: (a-d) Image width 8~$\upmu$m, (h,i) 800~nm. }
    \label{F:RefChip}
\end{figure}

\section{Evaluation of transferred samples}
\label{S:Eval}
 
During the development and refinement of the transfer procedure, various samples have been transferred. In this section we first discuss a reference sample, i.e. a transfer of a SiN\textsubscript{x} membrane without nanomaterial. The transfer quality is then assessed for a 1D nanomaterial followed by the GO sample presented in Figure~\ref{F:GOtrans} and a monolayer of WS\textsubscript{2}. Finally, electrical measurements are presented and discussed.
 
\subsection{Reference sample preparation}
\label{Sref}

To study the impact of the sample preparation process and to establish a baseline for electrical measurements, we prepared a reference chip using a carbon-coated SiN\textsubscript{x} membrane only, i.e., without any nanomaterial. \textbf{Figure~\ref{F:RefChip}} shows SEM images (a-d), electrical measurements (e-g) as well as (S)TEM and EDX analysis (h-j) of the chip at different stages of the process.  In the preparation of this reference sample, step 10 (contacting of membrane to chip contacts by FIBID) and step 11 (removal of needle) were switched to check if the attractive forces between the membrane and the contact pads are strong enough for the membrane to stay in contact even without prior fixation by FIBID. Figure~\ref{F:RefChip}c shows that this is indeed the case as the membrane remains in its previous position even after retraction of the micro needle.  The FIBID deposition to contact the nanomaterial and membrane with the pads (Figure~\ref{F:RefChip}d) is however necessary to ensure a mechanical fixation before venting the SEM-FIB instrument and is also required to guarantee a good electrical contact. 

The I-V curves displayed in Figure~\ref{F:RefChip}e were acquired at four stages of the preparation process: (i) of the raw chip (solid line), (ii) after milling the hole between the contacts (dashed line), (iii) after transferring the membrane (dotted line) and (iv) after milling away the SiN\textsubscript{x} membrane (blue dashed line). When comparing the curves before and after milling of the hole between the contacts, an increase in current is observed, which amounts to almost one order of magnitude for chip 1 and is smaller for a second reference chip 2 (orange lines). This increase in current and related decrease in resistance (Figure~\ref{F:RefChip}g), is related to Ga incorporation in the SiN\textsubscript{x} membrane. The noisy I-V curve of the untreated chip contains a non-linear component (Figure~\ref{F:RefChip}f), which is expected for insulating SiN\textsubscript{x}. The curve is linear after milling of the hole induced by the Ga implantation at the edge of the hole.  A straightforward way to reduce the contribution of the chip to the measured resistance is to mill a longer hole, which increases the length of the current path and thus reduces its contribution to the conductivity. 

A jump in current of over two orders of magnitude occurs when the C-coated SiN\textsubscript{x} membrane is transferred to the chip (Figure~\ref{F:RefChip}e) and the resistance decreases from $\approx$80 MOhm to 400~kOhm (Figure~\ref{F:RefChip}g). This shows that one can assume that only a negligible amount of the applied current will pass through the chip once a material with metallic or semiconducting behavior and a resistance below 1 MOhm is placed on the chip. The I-V curve acquired after removal of the SiN\textsubscript{x} membrane (light blue dashed line Figure~\ref{F:RefChip}d)  is identical to the one acquired before the transfer, indicating that redeposition  on the border of the hole  is negligible and is not affecting the electrical properties of the chip.

 The (S)TEM images and EDX analysis (Figure~\ref{F:RefChip}h-j) show that contamination is strongly minimized via this preparation process. While the presence of Pt is clearly observed directly at the membrane edge located close to the area where the needle was contacted from the bright edge in Figure~\ref{F:RefChip}i and the blue spectrum in Figure~\ref{F:RefChip}j, a minimum amount of Pt contamination is seen in the center (yellow spectrum in Figure~\ref{F:RefChip}j) and at the opposite edge (red spectrum). Ga is found on both edges and the central area in a low amount, implanted during the cutting processes of the membrane (Steps 5 and 12). 
 
\subsection{1D materials}
 
\begin{figure}[t]
    \centering
    \includegraphics[width=0.75\linewidth]{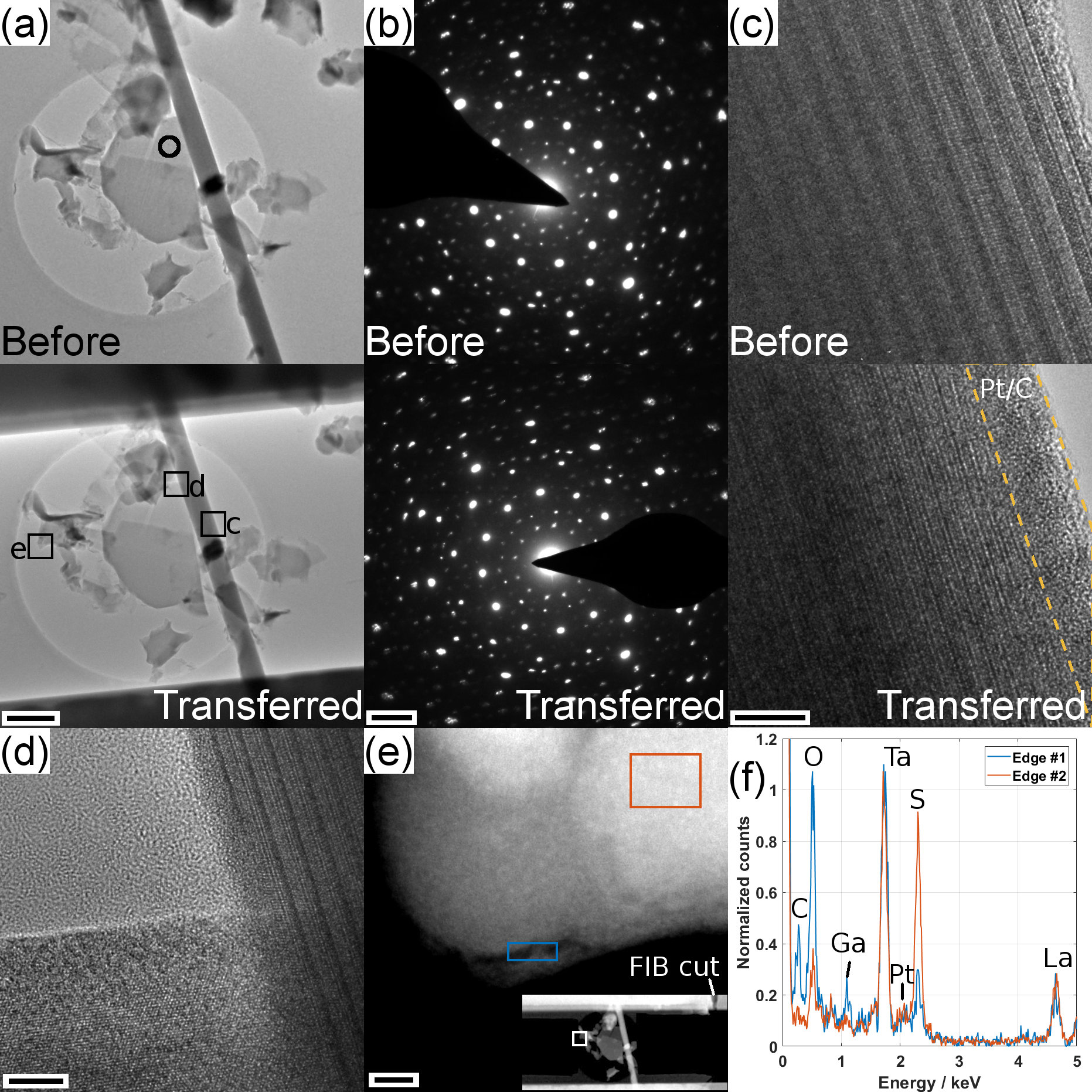}
    \caption{Evaluation of the transfer of a NT and flake made of the MLC LaS-TaS\textsubscript{2}. (a-c) Comparison of (a) low-mag TEM images, (b) SAED patterns and (c) HRTEM images before and after the sample transfer showing negligible damage and minimum contamination. Black circle in (a) marks position for SAED acquisition. (d) HRTEM image of the area marked in (b) shows the preservation of the sample structure.  (e) HAADF-STEM image of the edge of a flake close to the SiN\textsubscript{x} border marked in the inset HAADF-STEM image (width 8~$\upmu$m) and in (a). (f) STEM-EDX comparison of two areas at the edge marked in (e).   Scale bars are (a) 500~nm, (b) 2~nm\textsuperscript{-1}, (c-d) 5~nm and (e) 10~nm. }
    \label{F:FigEva}
\end{figure}
 
 As an example for a 1D nanomaterial, we discuss the LaS-TaS\textsubscript{2} NT and flake sample, whose transfer is described in Figure~S1. \textbf{Figure~\ref{F:FigEva}}a-d shows a first evaluation of the transfer after step 11, prior to  the removal of the SiN\textsubscript{x} in the hole area.  In the TEM images taken at lower magnification (Figure \ref{F:FigEva}a), no change induced by the transfer process is observable.  In the lower image of Figure~\ref{F:FigEva}a, the edges of the hole in the in-situ chip are visible at the top and bottom of the TEM image. In Figure~\ref{F:FigEva}b, the fine structures of the rich selected-area electron diffraction (SAED) pattern acquired of the MLC flake as marked in (a) are preserved after the transfer. 

When looking at the right border of the transferred NT, a thin (less than 5~nm) amorphous deposit, which was not present before the transfer process, can be observed (marked  in the lower image of Figure~\ref{F:FigEva}c). This amorphous contamination is attributed to the deposition of Pt-C precursor gases by secondary electrons or ions created during the FIB milling of the membrane after contacting the micro needle to the sample and after contacting of the sample to the in-situ chip. The deposit is only observed on the right side of the NT as the location of FIB milling and needle contact  lies to the right side of the NT (c.f. Figure~\ref{F:FigEva}e and Figure~S1) and then, the generated secondary electrons or ions cannot easily reach the left side of the NT. The left part of the NT and the flake are therefore perfectly preserved after transfer (Figure~\ref{F:FigEva}d).

Figure~\ref{F:FigEva}e shows a HAADF-STEM image of the sample with removed SiN\textsubscript{x} membrane after performing the final step 12 of the transfer process. The image was acquired from the edge of a flake located closely to the edge of the hole in the SiN\textsubscript{x} membrane as marked in Figure~\ref{F:FigEva}a and in the inset image in (e). Two EDX spectra from the very edge (blue frame) and from a thicker part located less then 100~nm away from the edge (orange frame) were acquired, which are depicted in Figure~\ref{F:FigEva}f. For the sake of comparison, the spectra were normalized with respect to the Ta-M edge at 1.7~keV. The main metallic constituents of the MLC material (La, Ta) are observed in similar amounts in both spectra. Sulfur is strongly reduced at the edge of the flake (blue line) and, in contrast, contamination elements are found at the edge of the flake (C, O, Ga). The Ga can be linked to an implantation during the cutting process of the membrane, which is however spatially confined to the very edge of the sample. Additional damage or contamination of the actual NT and flake during the final milling step were not observed.

Further results obtained on 1D nanomaterials can be found in the SI. As an example for correlative microscopy, Figure~S2 shows a Raman spectrum obtained from the MLC NT and flake sample with modes corresponding to both the LaS and the TaS\textsubscript{2} subsystems in agreement with literature \cite{Hettler.2020,Radovsky.2016}.  To emphasize the reproducibility of the proposed transfer process, Figures~S3 and S4 depict the transfer evaluation using TEM images, SAED patterns and EDX analysis of a second LaS-TaS\textsubscript{2} NT and a WS\textsubscript{2} NT \cite{Sreedhara.2022}, which both are preserved in their original state. 

\subsection{Graphene oxide}
 
\begin{figure}[t]
    \centering
    \includegraphics[width=0.8\linewidth]{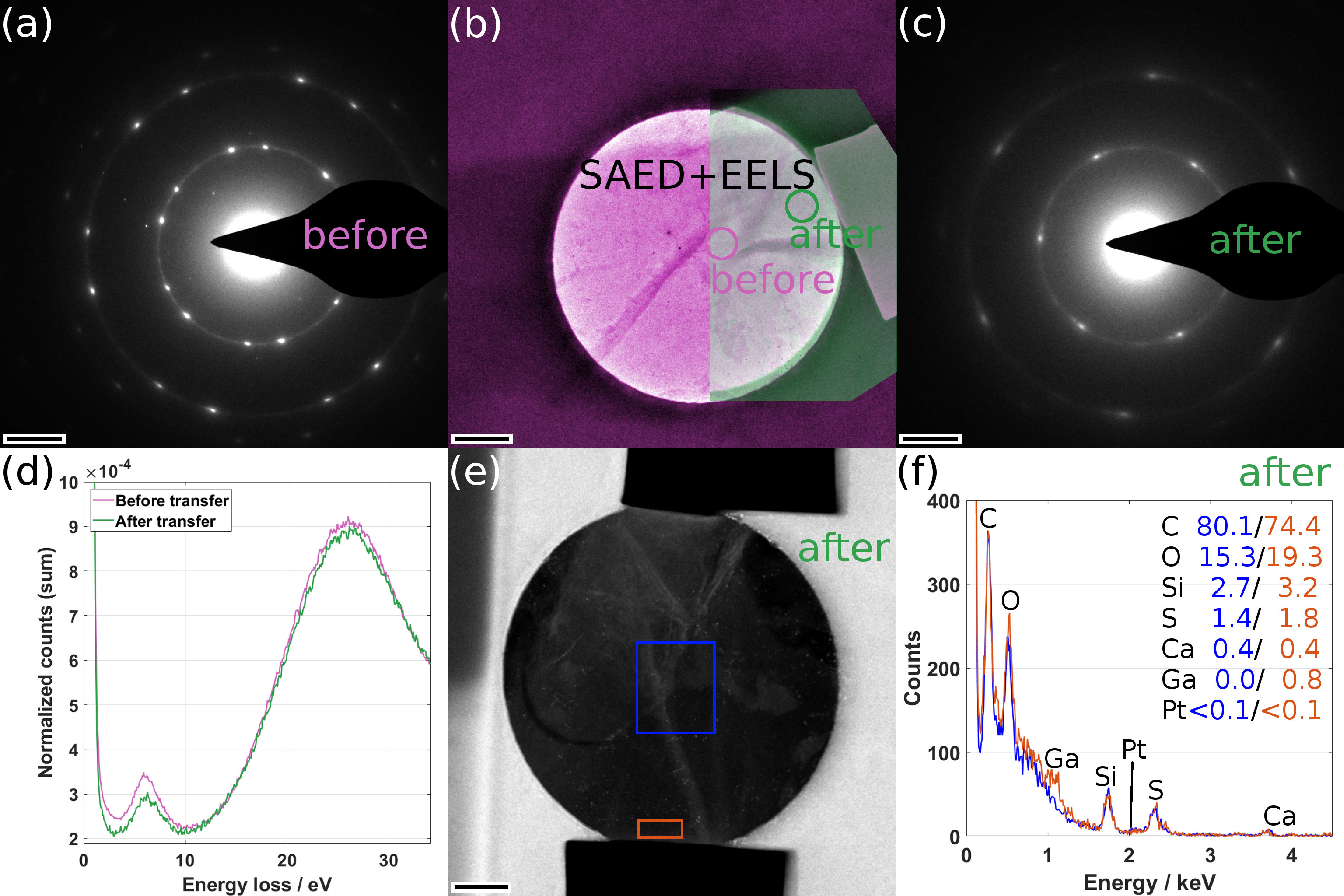}
    \caption{ Evaluation of transfer quality for the GO specimen whose transfer is described in Figure~\ref{F:GOtrans}. (a) SAED pattern of GO before transfer showing two sets of graphene lattices. (b) Blend of two TEM images before (light pink) and after (green) transfer to the in-situ chip. Marked positions correspond to both SAED pattern and EELS acquisition locations. (c) SAED pattern after transfer revealing a single graphene lattice (taken from another site of the GO film). (d) Comparison of low-loss EEL spectra before and after transfer. (e) STEM image after transfer. (f) EDX spectra taken from areas marked in blue and orange in (e) showing mainly carbon and oxygen and some typical GO contamination Ca, Si, S in low amounts. Ga implantation is observed at the edge of the film and no Pt can be detected. (a,c) 3~nm\textsuperscript{-1}, (b) 300~nm, (e) 400~nm.  }
    \label{F:FigEva2}
\end{figure}

\textbf{Figure~\ref{F:FigEva2}} describes the evaluation of the GO specimen before and after the transfer described in Figure~\ref{F:GOtrans} as an example for a 2D material. The graphene backbone of the heterogeneous GO film, made up of several individual flakes piled up on top of each other is clearly preserved as shown by the SAED patterns acquired from the positions marked in Figure~\ref{F:FigEva2}b before (Figure~\ref{F:FigEva2}a) and after (Figure~\ref{F:FigEva2}c) the transfer. The superposition of two TEM images before and after the transfer (Figure~\ref{F:FigEva2}b) shows that the morphology of the film, i.e. the folded up areas, is not altered by the transfer. TEM-EELS analysis obtained from the identical sites as the SAED patterns reveal both the surface (6~eV) and bulk (25~eV) plasmons of the GO. STEM-EDX analysis was performed to check the chemical composition of the flake after transfer. Two spectra acquired from the center and the edge of the GO film show that the predominant composition of C and O and even minor natural contamination (Si, S, Ca) of the sample is maintained during the transfer process. Pt contamination could not be found by EDX analysis after the transfer ($<$~0.1~at\%). However, slight contamination from Ga implantation indeed is observed at the edge of the film where the FIB was used to remove the SiN\textsubscript{x} membrane. In the case of this GO sample, the Ga implantation does not lead to a noticeable modification of its structure and properties (c.f. section~\ref{S:elec}).

\newpage
\subsection{Monolayer WS\textsubscript{2}}
 
\begin{figure}[h]
    \centering
    \includegraphics[width=0.8\linewidth]{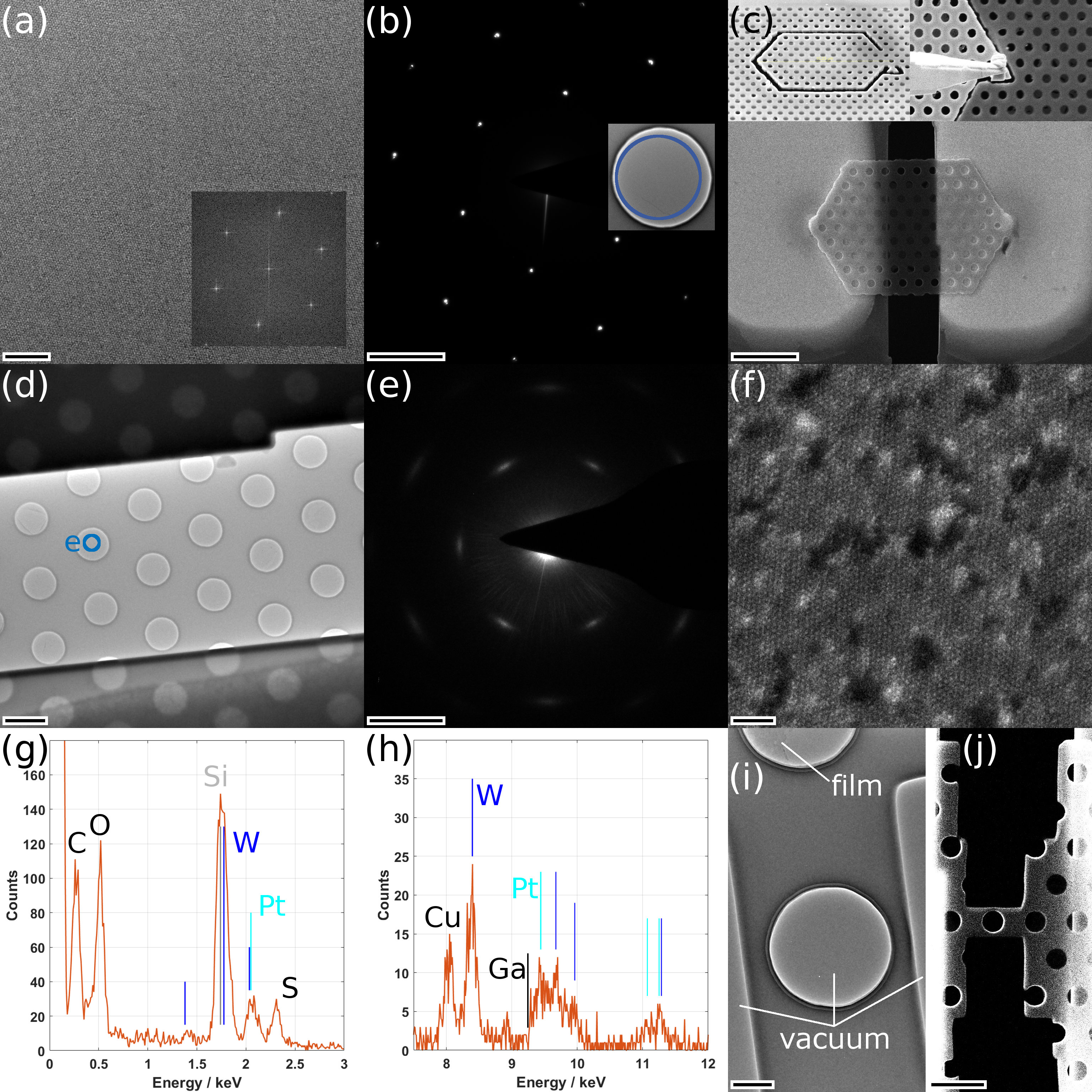}
    \caption{Analysis of WS\textsubscript{2} monolayer sample transfer. (a) Unfiltered HRTEM image with inset power spectrum and (b) SAED pattern of the WS\textsubscript{2} monolayer transferred to the SiN\textsubscript{x} membrane via a polystyrene-based floating approach \cite{Gurarslan.2014}. (c) Three SEM images of the support-based sample transfer to an in-situ chip. (d) TEM image of the transferred sample to the chip. (e) SAED pattern taken from the area marked in (d). (f) HRSTEM image obtained from the same area with an electron energy of 120~keV revealing the WS\textsubscript{2} lattice. (g,h) STEM-EDX spectrum acquired from the film area with (g) the low- and (h) the high-energy range, revealing the presence of C, O, Si and Pt contaminants, a fluorescence signal of Cu as well as W and S. No Ga can be detected. (i) TEM and (j) SEM image of the sample after the final SiN\textsubscript{x} milling showing the absence of the film in the central area. Scale bars are (a) 3~nm (inset FFT width =~10~nm\textsuperscript{-1}), (b) 3~nm\textsuperscript{-1} (inset TEM image width =~700~nm), (c) 3 $\upmu$m, inset image widths 8.3 (left) and 13.8 $\upmu$m (right), (d) 400~nm, (e) 3~nm\textsuperscript{-1}, (f) 2~nm, (i) 100~nm and (j) 1~$\upmu$m.}
    \label{F:FigWS2}
\end{figure}

In the shown example sample transfer processes of nanomaterials, damage of the sample could not be observed and contamination is limited to a very thin amorphous Pt/C layer and minor Ga implantation at the edge of the 2D material GO. For the shown larger 1D or 2D inorganic nanomaterials, this contamination is small, allowing to focus on the materials' properties in in-situ electrical TEM studies. To study the limits of the proposed method, we tested it to transfer a monolayer WS\textsubscript{2} sample epitaxially grown on sapphire \cite{Cohen.2023}. In a first step, the WS\textsubscript{2} was transferred from the sapphire substrate to a SiN\textsubscript{x} TEM grid using a polystyrene-based floating approach (Figure~S5)\cite{Gurarslan.2014}. Secondly, the monolayer, sustained by the SiN\textsubscript{x} holey membrane, was transferred to an in-situ chip by the method proposed in this work, at first leaving out the final step 12 of removing the SiN\textsubscript{x} membrane. The results described in \textbf{Figure~\ref{F:FigWS2}} were obtained with an operating voltage of 5~kV for the Ga ion beam for all the process steps. Figure~\ref{F:FigWS2}a and b show a HRTEM image and a SAED pattern from the monolayer on the SiN\textsubscript{x} TEM grid after the first transfer. Figure~\ref{F:FigWS2}c gives an overview of the FIB-based transfer by three SEM images. The TEM image of the transferred sample in Figure~\ref{F:FigWS2}d reveals that, in most of the holes in the central area of the holey membrane, the film remained intact. A comparison of the SAED pattern before (b) and after (e) the transfer (not obtained on the identical sample area) indicates that, although WS\textsubscript{2} remains the largest contribution, the reflections have smeared out considerably and have also lost in intensity. This is confirmed by the HRSTEM image in Figure~\ref{F:FigWS2}f, in which the WS\textsubscript{2} lattice is found in the major part of the area with dark amorphous parts and brighter particles.

An average EDX spectrum of the WS\textsubscript{2} sample obtained from various holes with intact film is shown in Figure~\ref{F:FigWS2}d and reveals the presence of W and S. When quantifying the individual spectra from different holes, the ratio between W and S is found to vary, being close to 1:2 in the central areas and with reduced S content in the outer areas. A significant contribution is Pt, which, however, is found with lesser intensity when compared to W, meaning less than one monolayer. As a part of the Pt signal might stem from fluorescence in the Pt contacts of the microchip, the Pt contamination might even be lower. The presence of the Cu peak, originating from the sample holder, indicates that fluorescence is indeed a non-negligible contribution to the spectrum. No Ga peak can be found showing that implantation and redeposition are, prior to the final step of milling the SiN\textsubscript{x}, minimum. Additional contamination is given by C, O and Si. An EELS comparison of the sample before and after the transfer shows that Si (nitride) contamination was already present before the transfer and is even slightly reduced afterwards, while C and O are enhanced (Figure~S6).

To again demonstrate the reproducibility of the approach, Figures~S7 and S8 depict two additional sample transfers of the WS\textsubscript{2} monolayer sample, where operating voltages of 5 and 16~ kV (Figure~S7) and 16~kV only (Figure~S8) were employed. The examples confirm that lowering the voltage of the Ga ions during and directly after the FIBID processes (steps 6 \& 7 and 10 \& 11)  to 5~kV decreases the damage and contamination inflicted on the WS\textsubscript{2} monolayer, showing that this step is crucial to minimize damage and contamination. In contrast, the operating voltage used for the cutting of the membrane (step 5)  is not a critical parameter as results shown in Figure~\ref{F:FigWS2} and S7 are similar, \textit{i.e.}, both show a reduced damage density in the WS\textsubscript{2} monolayer when comparing to Figure~S8. 

The results on the WS\textsubscript{2} monolayer presented so far were obtained on samples before performing the final step 12 of milling away the SiN\textsubscript{x} membrane. Figure~\ref{F:FigWS2}i and j show two TEM and SEM images of the sample after this final step, revealing that the film is completely damaged even though it has not been in direct contact with the primary Ga\textsuperscript{+} beam. The generated secondary ions thus are sufficient to induce strong damage within the range of a few hundred nm.

In summary, Pt contamination related to FIBID can be estimated to less than a monolayer. Damage by the Ga\textsuperscript{+} beam is as well less than a monolayer if the distance between the primary beam and the sample can be kept larger than aproximately 1~$\upmu$m. The dominant damaging process is the reduction of WS\textsubscript{2} (removal of S), which initially leads to a local amorphisation and finally to a rupture of the film. With improvements in the design of the SiN\textsubscript{x} membrane and the use of alternative ion beams, e.g. He or Ar, this damage can probably be reduced even further.

\clearpage
\newpage

\subsection{Electrical measurements}
\label{S:elec}
\begin{figure}[t]
    \centering
    \includegraphics[width=0.6\linewidth]{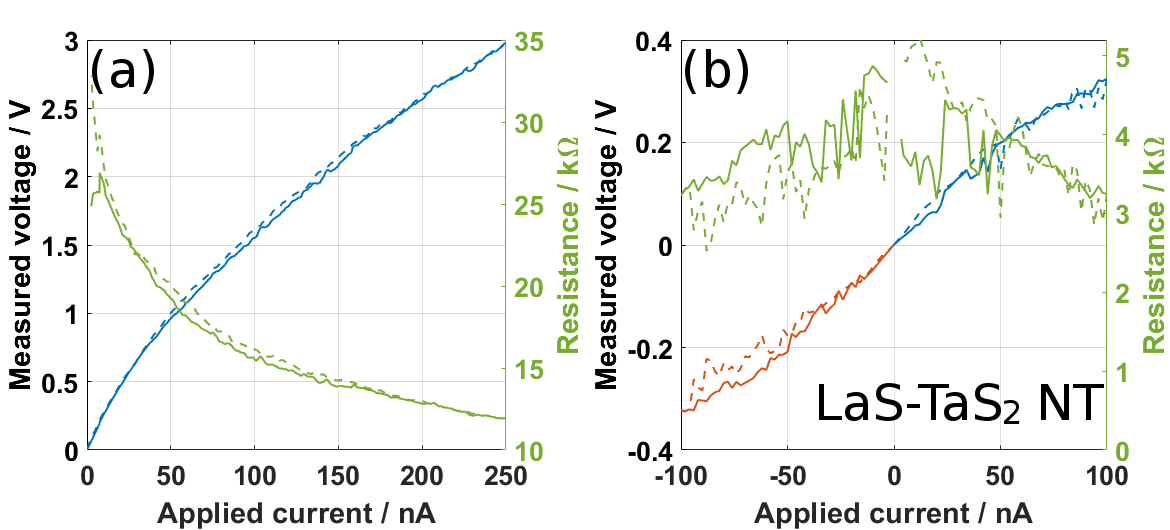}
    \caption{Electrical characterization of the (a) GO and (b) LaS-TaS\textsubscript{2} NT and flake samples, respectively. (a) Application of a double current sweep up to 250~nA with measured voltage and calculated resistance. (b) Application of two double current sweeps up to $\pm$100~nA with measured voltage and calculated resistance.}
    \label{F:Elec}
\end{figure}

\textbf{Figure~\ref{F:Elec}} shows the electrical characterization of the GO (Figures~\ref{F:GOtrans} and \ref{F:FigEva2}) and of the MLC NT and flake sample (Figures~S1 and \ref{F:FigEva}).  An electrical current ramp up to 250~nA (blue line in Figure~\ref{F:Elec}a) was applied to the GO sample. The ramp was conducted as double sweep, i.e. up (solid lines) and down (dashed lines), and only minor differences between both directions can be observed. A minor step at approximately 150~nA indicates a small change in the samples' conductivity. At low currents, a gap between up and down is observed in the resistance curve, which could be linked to the accuracy of the measurement device, which has a step in the measurement range at 10~nA. Apart from that, the V-I curve is smooth, indicating an Ohmic contact between GO and contact pad. The evolution is logarithmic, which corresponds to a decrease in resistance from 30 to 12~k$\Upomega$ (green curve) indicating the semiconducting nature of the material. 

Figure~\ref{F:Elec}b shows two double current ramps to 100~nA (blue lines) and -100~nA (red lines) applied to the MLC sample. The onset at 0~nA is linear and the curves do not exhibit steps indicating a good electrical contact with the sample. Again, the curves follow a logarithmic evolution due to the semiconducting nature of the MLC material and the resistance shows a slight decrease (green lines). The measurements however have a higher noise when comparing with the GO sample, which could be attributed to a lower quality of the Pt/C welding point as it has to bridge the insulating SiN\textsubscript{x} membrane, which for this sample is located between the contact pad and the NT.

The sample conductivity can be be calculated from the resistance if the samples' dimensions are known. In case of the GO sample, the width is measured from SEM and TEM images to 2.5~$\upmu$m and the average thickness can be estimated from low-loss EELS analysis to 70~nm. Assuming a good, areal contact between GO and the pad and neglecting the contribution from the Mo-coated SiN\textsubscript{x} membrane, the length is assumed as the distance between the edges of the pads to 5~$\upmu$m. With these values and a resistance of 30~k$\Upomega$, the conductivity calculates to 1~Sm\textsuperscript{-1}, which agrees with values from previous studies on a similar sample \cite{Hettler.2021}.

In case of the MLC NT and flake, the inhomogeneity of the sample only allows an estimation of the conductivity. As the NT has a rather large diameter (210~nm) in comparison to the flake thickness, we assume the NT as sole contribution to the conductivity, neglecting thus the flake. As for this sample the SiN\textsubscript{x} is between the NT and the pad, the length is given by the distance of 4.9~$\upmu$m between the Pt/C welding points at both ends of the NT (Figure~S1). With these values, the conductivity calculates to 8.8~Sm\textsuperscript{-1}, which is only slightly higher than the value obtained from SEM characterization of a similar NT \cite{Hettler.2020}, confirming the assumption that the NT represents the major contribution to the samples' conductivity. These analyses of the conductivity show that the sample geometry is of great importance when aiming for the determination of material properties from in-situ TEM measurements of individual nanomaterials. 

\section{Conclusion}

We have introduced a method facilitating and improving the transfer and contacting of individual nanomaterials to microchips with electrical contacts in a reproducible way, e.g.,  for electrical and/or thermal characterization by in-situ transmission electron microscopy (TEM). The benefit of using a SiN\textsubscript{x} membrane as sample support for a focused ion beam (FIB) based transfer of the nanomaterial is threefold: It firstly allows to move the necessary contact point between lift-out needle and specimen away from the area of interest to avoid damage and minimize contamination. Secondly, it provides the mechanical stability needed to transport the nanomaterial without bending, which on its own might not possess a sufficient mechanical rigidity. Finally, it guarantees a good attachment and subsequent contacting to the metal contacts of the in-situ chip due to electrostatic charging.

With the explained process, we prepared different 1D and 2D nanomaterials with minimum damage and contamination, which is the basis for a successful in-situ experiment. Remaining Pt contamination from focused ion beam induced deposition is found to be less than a monolayer. Ga implantation and redeposition can be completely avoided for 1D nanomaterials and is confined to the edges of 2D materials. Further improvements in the design of the SiN\textsubscript{x} grid and the use of alternative ion beams (He, Ne or Ar) could yield a completely artifact-free transfer process.

The method allows in-situ experiments of different nature, including an electrical characterization linked to structure and composition at the atomic level and failure analysis upon application of electrical currents. Moreover, studies of electron-beam induced current (EBIC) or the formation of electrical current paths are possible. These studies can be combined with additional heating or possibly be performed in liquid or gas environment using specific sample holders. Moreover, further correlative microscopy and spectroscopy techniques, which allow the local analysis in the 1~$\upmu$m range, can be performed before and after the in-situ TEM experiments.

In addition to in-situ TEM studies, the presented transfer can be utilized to transfer nanomaterials to any other microchip that possesses electrical contacts (with or without a hole in between) and where it is not possible to synthesize the nanomaterial directly on the chip.

\section{Experimental Section}

\label{S:MatMet}

\threesubsection{Materials}

Three TEM grids were investigated as possible supports for the in-situ transfer process. (1) Thicker carbon films (C-Flat, 40 nm), (2) UltrAu gold films (Quantifoil) and (3) silicon nitride (SiN\textsubscript{x}) films (PELCO) with regular arrays of holes were tested.  As discussed in section 2.3, only the SiN\textsubscript{x} grids were used for this work. These  consist in SiN\textsubscript{x} membranes with 200 nm thickness and an array of holes either with a diameter of 2.5 $\upmu$m or with multiple diameters between 70 and 1.25~$\upmu$m. As SiN\textsubscript{x} is insulating, a conductive coating is necessary to avoid charging in TEM and SEM, which has to be chosen depending on the composition of the nanomaterial and on the desired in-situ application. Experiments were performed with Pd and Mo coatings by sputter deposition (Leica EM ACE200 and AJA Orion 5 UHV) or C coatings by carbon-thread evaporation (Leica EM ACE200) with thicknesses below 20~nm. 

For demonstration of the preparation process, we have selected various nanomaterials: NTs and flakes of the misfit-layered compounds (MLCs) LaS-TaS\textsubscript{2} \cite{Lajaunie.2018,Hettler.2020}, WS\textsubscript{2} in the form of nanotubes \cite{Sreedhara.2022} and as epitaxially grown monolayer\cite{Cohen.2023} and graphene oxide (GO), synthesized by oxidation of graphite flakes using a modified Hummers method \cite{GO_SynthVictorRoman.2020}.

\threesubsection{Instrumentation and methods}

The FIB transfer was performed with a Helios 650 dual-beam instrument (Thermo Fisher Scientific), equipped with a Ga-ion source, an Omniprobe needle and a Pt-based precursor gas for focused electron beam (FEBID) or focused ion beam induced deposition (FIBID) of contacts. Employed parameters are detailed in the respective section. Both TEM grid and in-situ chip were mounted on two stripes of conductive Cu tape and two stripes of conductive C tape were used to improve fixation and guarantee good grounding of the metal contacts of the in-situ chip located on its top side.

The TEM experiments were conducted in two aberration-corrected Titan microscopes (Thermo Fisher Scientific). TEM, EELS in TEM mode (parallel illumination and collection angle of 11.9~mrad) and selected-area electron diffraction (SAED) was performed in the image-corrected instrument equipped with a Gatan Image Filter (GIF) Tridiem and operated at an electron energy of 300~keV, which was decreased to 80~keV for the GO sample. Scanning (S)TEM imaging, EDX and SAED were conducted in the probe-corrected microscope with a high-brightness gun (X-FEG) operated at 300 keV (120~keV for the WS\textsubscript{2} monolayer sample and 80~keV for GO) and an Oxford Instruments Ultim X-MaxN 100TLE detector for EDX measurements. EDX data was quantified using the Aztec software (Oxford Instruments) with theoretical k-factors.

In-situ experiments were performed with a DENSsolutions Wildfire (4 pins) sample holder. A Keithley Instruments 2450 SourceMeter (Tektronix) was used as current/voltage supply and measuring device for electrical characterization. For the in-situ electrical characterization, we designed and fabricated chips with a heating device and contacts for two-probe electrical characterization (Sketch and SEM image in Figure~S9 (SI)): Starting from a 4" (100) Si wafer with 380 $\upmu$m thickness, coated with 1 $\upmu$m of low-stress SiN\textsubscript{x} on both sides, two optical lithography steps were employed. The first one using a Ti35 ESX neg. photoresist (Microchemicals) to define windows on the back side:

\begin{enumerate}
    \item Spin coating at 3000 rounds per minute (RPM)
    \item Bake for 3~min at 100 \degree C (hot plate)
    \item Exposure with a dose of 175 mJ/cm\textsuperscript{2}
    \item 15 min waiting time
    \item Postbake for 3~min at 150\degree C (hot plate)
    \item Development in 1:1 H\textsubscript{2}O:AZ Developer (Merck) for 2 min
    \item Reactive ion etching (20 sccm SF6, 0.19 mbar, 200 W, ~15 min)
    \item Cleaning in acetone
\end{enumerate} 

The second lithography process was performed with the photoresist AZ 5214E to define the metal contacts on the top side:
\begin{enumerate}
    \item Spin coating at 6000 rounds per minute (RPM)
    \item 30 s waiting time
    \item Bake for 50~s at 110 \degree C (hot plate)
    \item Exposure with a dose of 26 mJ/cm\textsuperscript{2}
    \item Postbake for 60~s at 120\degree C (hot plate)
    \item Flood exposure with a dose of 390~mJ/cm\textsuperscript{2}
    \item Development in AZ 726 MiF (Merck) for 30~s
    \item Sputtering process of 5 nm Ti + 150 nm Pt (AJA Orion 5 UHV)
    \item Lift-off process in acetone assisted by ultrasound
\end{enumerate} 

Finally, the membrane is created by defined wet etching of the (100) Si in a KOH bath at 80\degree C using a holder with two-sided O-ring sealing for protection of the Pt contacts on the front side. The wafer is cut into individual chips with a wafer saw. 

Raman spectra were acquired with a confocal Raman Alpha 300 M+ (WiTec) with a 633~nm laser operated at 0.5 mW power and a 100x objective with numerical aperture of 0.9. The spectrometer was operated with 1800 grooves/mm grating.

\medskip
\textbf{Supporting Information} \par 
Supporting Information is available from the authors.

\medskip
\textbf{Acknowledgements} \par 

The authors acknowledge funding from the European Union’s Horizon 2020 research and innovation programme under the Marie Sklodowska-Curie grant agreement No 889546, by the Spanish MICN (PID2019-104739GB-100/AEI/10.13039/501100011033) and from the European Union H2020 programs “ESTEEM3” (Grant number 823717). The microscopy works have been conducted in the Laboratorio de Microscopias Avanzadas (LMA) at Universidad de Zaragoza. Sample courtesy from MB Sreedhara and R. Tenne (WS\textsubscript{2} and MLC NTs and flakes, Weizmann Institute of Science, Israel), A. Cohen and A. Ismach (WS\textsubscript{2} monolayer, Tel Aviv University, Israel) and W. Maser and A. Benito (GO, Instituto de Carboquimica, CSIC, Zaragoza) is acknowledged. We thank I. Echañiz (INMA) for support with acquisition of Raman spectra.

\bibliographystyle{MSP}
\bibliography{Biblio}

\begin{thebibliography}{10}
\providecommand{\url}[1]{\texttt{#1}}
\providecommand{\urlprefix}{URL }

\bibitem{Yao.2000}
Z.~Yao, C.~L. Kane, C.~Dekker,
\newblock \emph{Physical Review Letters} \textbf{2000}, \emph{84}, 13 2941.

\bibitem{Huang.2015}
J.-W. Huang, C.~Pan, S.~Tran, B.~Cheng, K.~Watanabe, T.~Taniguchi, C.~N. Lau,
  M.~Bockrath,
\newblock \emph{Nano Letters} \textbf{2015}, \emph{15}, 10 6836.

\bibitem{Brunbauer.2016}
F.~M. Brunbauer, E.~Bertagnolli, J.~Majer, A.~Lugstein,
\newblock \emph{Nanotechnology} \textbf{2016}, \emph{27}, 38 385704.

\bibitem{Lin.2018}
Z.~Lin, R.~Zhan, L.~Li, H.~Liu, S.~Jia, H.~Chen, S.~Tang, J.~She, S.~Deng,
  N.~Xu, J.~Chen,
\newblock \emph{RSC Advances} \textbf{2018}, \emph{8}, 4 2188.

\bibitem{Empante.2019}
T.~A. Empante, A.~Martinez, M.~Wurch, Y.~Zhu, A.~K. Geremew, K.~Yamaguchi,
  M.~Isarraraz, S.~Rumyantsev, E.~J. Reed, A.~A. Balandin, L.~Bartels,
\newblock \emph{Nano Letters} \textbf{2019}, \emph{19}, 7 4355.

\bibitem{Roy.2022}
K.~S. Roy, S.~Hettler, R.~Arenal, L.~S. Panchakarla,
\newblock \emph{Materials Horizons} \textbf{2022}, \emph{9}, 8 2115.

\bibitem{Sato.2017}
Y.~Sato, T.~Gondo, H.~Miyazaki, R.~Teranishi, K.~Kaneko,
\newblock \emph{Applied Physics Letters} \textbf{2017}, \emph{111}, 6 062904.

\bibitem{Zhang.2017}
Q.~Zhang, X.~He, J.~Shi, N.~Lu, H.~Li, Q.~Yu, Z.~Zhang, L.-Q. Chen, B.~Morris,
  Q.~Xu, P.~Yu, L.~Gu, K.~Jin, C.-W. Nan,
\newblock \emph{Nature Communications} \textbf{2017}, \emph{8}, 1 4791.

\bibitem{MolinaLuna.2018}
L.~Molina-Luna, S.~Wang, Y.~Pivak, A.~Zintler, H.~H. P{\'e}rez-Garza, R.~G.
  Spruit, Q.~Xu, M.~Yi, B.-X. Xu, M.~Acosta,
\newblock \emph{Nature Communications} \textbf{2018}, \emph{9}, 1 682.

\bibitem{Ishida.2020}
T.~Ishida, H.~Hiroshima, K.~Higuchi, M.~Tomita, K.~Saitoh, T.~Tanji,
\newblock \emph{Surface and Interface Analysis} \textbf{2020}, \emph{52}, 9
  584.

\bibitem{Arita.2020}
M.~Arita, A.~Tsurumaki-Fukuchi, Y.~Takahashi,
\newblock \emph{Japanese Journal of Applied Physics} \textbf{2020}, \emph{59},
  SG SG0803.

\bibitem{Luong.2020}
M.~A. Luong, E.~Robin, N.~Pauc, P.~Gentile, M.~Sistani, A.~Lugstein, M.~Spies,
  B.~Fernandez, M.~I. {den Hertog},
\newblock \emph{ACS Applied Nano Materials} \textbf{2020}, \emph{3}, 2 1891.

\bibitem{Nukala.2021}
P.~Nukala, M.~Ahmadi, Y.~Wei, S.~de~Graaf, E.~Stylianidis, T.~Chakrabortty,
  S.~Matzen, H.~W. Zandbergen, A.~Bj{\"o}rling, D.~Mannix, D.~Carbone, B.~Kooi,
  B.~Noheda,
\newblock \emph{Science} \textbf{2021}, \emph{372}, 6542 630.

\bibitem{Gorji.2020}
S.~Gorji, A.~Kashiwar, L.~S. Mantha, R.~Kruk, R.~Witte, P.~Marek, H.~Hahn,
  C.~K{\"u}bel, T.~Scherer,
\newblock \emph{Ultramicroscopy} \textbf{2020}, \emph{219} 113075.

\bibitem{denHertog.2012}
M.~I. {den Hertog}, F.~Gonz{\'a}lez-Posada, R.~Songmuang, J.~L. Rouviere,
  T.~Fournier, B.~Fernandez, E.~Monroy,
\newblock \emph{Nano Letters} \textbf{2012}, \emph{12}, 11 5691.

\bibitem{Kozlova.2013}
T.~Kozlova, M.~Rudneva, H.~W. Zandbergen,
\newblock \emph{Nanotechnology} \textbf{2013}, \emph{24}, 50 505708.

\bibitem{Hsueh.2023}
Y.-H. Hsueh, A.~Ranjan, L.-M. Lyu, K.-Y. Hsiao, Y.-C. Chang, M.-P. Lu, M.-Y.
  Lu,
\newblock \emph{Advanced Electronic Materials} \textbf{2023}, \emph{9}, 3 220.

\bibitem{MartialDuchamp.2014}
{Martial Duchamp}, {Qiang Xu}, {Rafal E Dunin-Borkowski},
\newblock \emph{Microscopy and Microanalysis} \textbf{2014}, \emph{20}, 6 1638.

\bibitem{Vijayan.2017}
S.~Vijayan, J.~R. Jinschek, S.~Kujawa, J.~Greiser, M.~Aindow,
\newblock \emph{Microscopy and Microanalysis} \textbf{2017}, \emph{23}, 4 708.

\bibitem{Xu.2011}
Z.~Xu, Y.~Bando, L.~Liu, W.~Wang, X.~Bai, D.~Golberg,
\newblock \emph{ACS Nano} \textbf{2011}, \emph{5}, 6 4401.

\bibitem{Hettler.2021}
S.~Hettler, D.~Sebastian, M.~Pelaez-Fernandez, A.~M. Benito, W.~K. Maser,
  R.~Arenal,
\newblock \emph{2D Materials} \textbf{2021}, \emph{8}, 3 031001.

\bibitem{Golberg.2006}
D.~Golberg, M.~Mitome, K.~Kurashima, C.~Y. Zhi, C.~C. Tang, Y.~Bando,
  O.~Lourie,
\newblock \emph{Applied Physics Letters} \textbf{2006}, \emph{88}, 12 1513.

\bibitem{Aslam.2011}
Z.~Aslam, R.~Nicholls, A.~{A. Koos}, V.~Nicolosi, N.~Grobert,
\newblock \emph{The Journal of Physical Chemistry C} \textbf{2011}, \emph{115},
  50 25019.

\bibitem{Arenal.2011}
R.~Arenal, M.-S. Wang, Z.~Xu, A.~Loiseau, D.~Golberg,
\newblock \emph{Nanotechnology} \textbf{2011}, \emph{22}, 26 265704.

\bibitem{Huang.2022}
X.~Huang, Y.~Tang, C.~K{\"u}bel, {Di Wang},
\newblock \emph{Microscopy and Microanalysis} \textbf{2022}, \emph{28}, 6 1981.

\bibitem{GO_SynthVictorRoman.2020}
S.~V{\'i}ctor-Rom{\'a}n, E.~Garc{\'i}a-Bordej{\'e}, J.~Hern{\'a}ndez-Ferrer,
  J.~M. Gonz{\'a}lez-Dom{\'i}nguez, A.~Ans{\'o}n-Casaos, A.~M.~T. Silva, W.~K.
  Maser, A.~M. Benito,
\newblock \emph{Catal. Today} \textbf{2020}, \emph{357} 350.

\bibitem{Lajaunie.2018}
L.~Lajaunie, G.~Radovsky, R.~Tenne, R.~Arenal,
\newblock \emph{Inorg. Chem.} \textbf{2018}, \emph{57}, 2 747.

\bibitem{Hettler.2020}
S.~Hettler, M.~B. Sreedhara, M.~Serra, S.~S. Sinha, R.~Popovitz-Biro,
  I.~Pinkas, A.~N. Enyashin, R.~Tenne, R.~Arenal,
\newblock \emph{ACS Nano} \textbf{2020}, \emph{14}, 5 5445.

\bibitem{Li.2021}
C.~Li, A.~P. Tardajos, D.~Wang, D.~Choukroun, K.~{van Daele}, T.~Breugelmans,
  S.~Bals,
\newblock \emph{Ultramicroscopy} \textbf{2021}, \emph{221} 113195.

\bibitem{Radovsky.2016}
G.~Radovsky, R.~Popovitz-Biro, T.~Lorenz, J.-O. Joswig, G.~Seifert, L.~Houben,
  R.~E. Dunin-Borkowski, R.~Tenne,
\newblock \emph{J. Mater. Chem. C} \textbf{2016}, \emph{4}, 1 89.

\bibitem{Sreedhara.2022}
M.~B. Sreedhara, Y.~Miroshnikov, K.~Zheng, L.~Houben, S.~Hettler, R.~Arenal,
  I.~Pinkas, S.~S. Sinha, I.~E. Castelli, R.~Tenne,
\newblock \emph{Journal of the American Chemical Society} \textbf{2022},
  \emph{144}, 23 10530.

\bibitem{Cohen.2023}
A.~Cohen, P.~K. Mohapatra, S.~Hettler, A.~Patsha, K.~V. L.~V. Narayanachari,
  P.~Shekhter, J.~Cavin, J.~M. Rondinelli, M.~Bedzyk, O.~Dieguez, R.~Arenal,
  A.~Ismach,
\newblock \emph{ACS nano} \textbf{2023}, \emph{17}, 6 5399.

\bibitem{Gurarslan.2014}
A.~Gurarslan, Y.~Yu, L.~Su, Y.~Yu, F.~Suarez, S.~Yao, Y.~Zhu, M.~Ozturk,
  Y.~Zhang, L.~Cao,
\newblock \emph{ACS nano} \textbf{2014}, \emph{8}, 11 11522.

\end{thebibliography}

\end{document}